\documentclass[prl,twocolumn,showpacs,showkeys]{revtex4}

\usepackage{graphicx}
\usepackage{dcolumn}
\usepackage{bm}
\usepackage{amssymb}

\begin{document}
\title{The  paramagnetic photon. Absence of perpendicular component and decay in large fields}
\author{ H. P\'erez Rojas and E. Rodriguez Querts}
\affiliation{Instituto de Cibernetica, Matematica y Fisica, Calle E
309, Vedado, Ciudad Habana, Cuba.\\}

\begin{abstract}
Previous results from the authors concerning the arising a tiny
photon anomalous paramagnetic moment $\mu_{\gamma}$ due to its
interaction with a magnetized virtual electron-positron background
are complemented and discussed. It is argued that such magnetic
moment it cannot be a linear function of the angular momentum and
that there is no room for the existence of an hypothetical
perpendicular component, as recently claimed in the literature. It
is discussed that in the region beyond the first threshold, where
photons may decay in electron-positron pairs, the photon magnetic
moment cannot be defined independently of the magnetic moment of the
created pairs. It is shown that for magnetic fields large enough,
the vacuum becomes unstable and decays also in electron-positron
pairs.
\end{abstract}



\maketitle

 \section{Introduction}
We have shown in \cite{EliPRD} that in analogy to the electron
anomalous magnetic moment $\mu^\prime=\alpha\mu_B/2\pi$ (being
$\mu_B=e/2m$ the Bohr magneton) shown by Schwinger \cite{Schwinger}
as due to their interaction with the virtual photon background
through its self-energy,  a similar effect exists for photons. Due
to the magnetic properties of the photon self-energy, a photon
anomalous magnetic moment $\mu_{\gamma}>0$ arises, which is
paramagnetic in the region of transparency (which is the region of
momentum space where the photon self-energy, and in consequence, its
frequency $\omega$, is real). The photon magnetic moment vanishes
only when its momentum $\textbf{k}$ is parallel to the magnetic
field $\textbf{B}$.

 The photon magnetic properties are due to the
 dependence of $\omega$   on $B$ expressed by the photon dispersion
 equation dependence on the
self-energy tensor $\Pi_{\mu\nu}(x,x^{\prime\prime}\vert A^{ext})$.
But let us recall some details about the electron-positron quantum
mechanics in a magnetic field. We assume some magnetic field defined
by the field invariants ${\cal F}=2B^2>0$, ${\cal G}=0$. In a given
coordinate system, constant uniform magnetic field $\textbf{B}$,
taken along $x_3$, produces a symmetry breaking of the space
symmetry. For electrons and positrons ($e^{\pm}$) physical
quantities are invariant only under rotations around $x_3$ or
displacements along it \cite{Johnson}. This means that the conserved
quantities, i.e., those commuting with the Hamiltonian operator, are
all parallel to $\textbf{B}$, as angular momentum and spin
components $J_3$,$L_3$,$s_3$ and the linear momentum $p_3$. By using
units $\hbar=c=1$, the energy eigenvalues for $e^{\pm}$ are
$E_{n,p_3}=\sqrt{p_3^2+m^2+ eB(2n+1+s_3)}$ where $s_3=\pm 1$ are
spin eigenvalues along $x_3$ and $n=0,1,2..$ are the Landau quantum
numbers. In other words, in presence of $B$, the transverse squared
energy $E_{n,p_3}^2-p_3^2$ is quantized by integer multiples of
$eB$. For the ground state $n=0$, $s=-1$, the integer is zero.
Quantum states degeneracy with regard spin is expressed by a term
$\alpha_n=2-\delta_{0n}$, whereas degeneracy with regard to orbit'
center coordinates leads to a factor $eB$, The quantity $1/eB$
characterizes the spread of the $e^{\pm}$ spinor wavefunctions in
the plane orthogonal to $B$. The magnetic moment operator $M$  is
defined as the quantum average of $M= -\partial H/\partial B$, where
$H$ is the Dirac Hamiltonian in the magnetic field $B$, and it is
not a constant of motion. But its time-dependent terms vanish after
quantum averaging and it leads to $\bar M=-\partial
E_{n,p_3}/\partial B$. Obviously, it must be understood as the
modulus of a vector parallel to $B$. There is no a linear relation
between $\bar M$ and $J_3$ as it exists in non-relativistic quantum
mechanics. There is also no room for conjecture the arising of a
magnetic moment component in a direction different from $B$.

 Due to the explicit symmetry breaking, the four momentum operator
acting on the vacuum state does not have a vanishing four-vector
eigenvalue, $P_{\mu}|0,B> \neq 0$. The components $P_{1,2}$ does not
commute with the Hamiltonian operator $H$, and are not conserved.
The $e^{\pm}$ quantum vacuum energy density is given by
$\Omega_{EH}=-eB\sum_{n=0}^\infty \alpha_n \int dp_3 E_{n,p_3}$,
After removing divergences it gives the well-known Euler-Heisenberg
expression $\Omega_{EH}=\frac{\alpha
B^2}{8\pi^2}\int_0^{\infty}e^{-B_c x/B}\left[\frac{coth x}{x}
-\frac{1}{x^2}-\frac{1}{3}\right]\frac{d x}{x}$ which is an even
function of $B$ and $B_c$, where $B_c=m^2/e\simeq 4.4\times
10^{13}$G is the Schwinger critical field. The magnetized vacuum is
paramagnetic ${\cal M}_V=-\partial \Omega_{EH}/\partial B>0$ and is
an odd function \cite{Elizabeth} of $B$. For $B<<B_c$ it is ${\cal
M}_V=\frac{2\alpha}{45 \pi}\frac{B^{3}}{B_c^2}$, where $\alpha$ the
fine structure constant. ${\cal M}_V$ is obviously be understood as
the modulus of a vector parallel to $\textbf{B}$.

The magnetic field $B$, is defined as such for a class of reference
frames moving parallel to $B$.  For frames moving in other
directions, for instance, perpendicular to $B$, the magnetic field
changes to $B'$ and an electric field $E$ arises, such that
$B'^2-E^2=B^2$. We do not intend to define an anomalous electric
moment in that case, since it would be of pure kinematical origin
(it can be eliminated by a Lorentz boost) and for that reason, it is
not interesting.

Most of this paper is based on a talk presented at the International
Workshop on Astronomy and Relativistic Astrophysics 2009 (IWARA09),
which was based on \cite{EliPRD}, as well as some new results. We
have enlarged the manuscript in some details, and discussed a new
reference related to the content of \cite{EliPRD}.

\section{The photon magnetic moment from  Shabad's dispersion equations}
In a similar way as the definition of the electron-positron magnetic
moment, for an observable photon we define the quantity
$\mu_{\gamma}=-\partial \omega/\partial B$, which is to be
understood also as the modulus of a vector along $\textbf{B}$. It
can be obtained analytically from the dispersion equations for the
photon in a magnetic field and it is generated by the photon
self-energy tensor dependence on $B$. There is no any reason to
impose a priori any linear relation between $\mu_{\gamma}$ and
classical photon spin, (or either electron-positron angular moment
operators) as has been made recently in the literature (See
\cite{Selym}. We will return later to this reference)

The diagonalization of the photon self-energy tensor leads to the
equations \cite{shabad1}
\begin{equation}
 \Pi_{\mu
\nu}a^{(i)}_{\nu}=\kappa_{i}a^{(i)}_{\mu},
\end{equation}
 having
three non vanishing eigenvalues
 and three eigenvectors for $i=1,2,3$, corresponding to three photon propagation modes. One additional eigenvector is
the photon four momentum vector $k_{\nu}$ whose eigenvalue is
$\kappa_{4}=0$. The first three eigenvectors
\begin{eqnarray} \nonumber
    a^{1}_\mu= k^2 F^2_{\mu \lambda}k^\lambda-k_\mu (kF^2 k),
     \\
    a^{2}_\mu=F^{*}_{\mu \lambda}k^\lambda,\hspace{.3cm} a^{3}_\mu=F_{\mu
\lambda}k^\lambda,
\end{eqnarray}
satisfy the four dimensional transversality condition
$a^{(1,2,3)}_{\mu}k_{\mu}=0$. Here $k_\mu$ is the photon
four-momentum, $F_{\mu \nu}=\partial_\mu A_\nu-\partial_\nu A_\mu$
and $F^{\mu \nu*}=\frac{1}{2}\epsilon^{\mu \nu \rho \kappa}F_{\rho
\kappa}$ are the external electromagnetic field tensor and its dual
pseudotensor, respectively.

In reference frames which are at rest or moving parallel to
$\textbf{B}$ we define $\textbf{n}_\perp=\textbf{k}_\perp/k_\perp$
and $\textbf{n}_\parallel=\textbf{k}_\parallel/ k_\parallel$ as the
transverse and parallel unit vectors respectively.

By  $a^{(i)}_\mu (x)$  as the electromagnetic four vector describing
these eigenmodes, its electric and magnetic fields ${\bf
e^{(i)}}=-\frac{\partial }{\partial x_0}\vec{a}^{(i)}-\frac{\partial
}{\partial {\bf x}}a^{(i)}_0$, ${\bf
h}^{(i)}=\nabla\times\vec{a}^{(i)}$ are obtained in \cite{shabad1}.
 It is easy to see
that the mode $i=3$ is a transverse plane polarized wave whose
electric unit vector is $\textbf{e}^{(3)}= (\textbf{n}_\perp\times
\textbf{n}_\parallel)$ orthogonal to the plane ($\textbf{B},
\textbf{k}$). For  $\bf k \perp \bf B$, $\textrm{a}^{(1)}_\mu$ is
longitudinal,  polarized along $\textbf{e}^{(1)}=\textbf{n}_\perp$
and it is a non physical mode, whereas $\textrm{a}^{(2)}_\mu$ is
transverse, since $\textbf{e}^{(2)}=\textbf{n}_\parallel$. Thus, in
that case modes $\textrm{a}^{(2,3)}_\mu$ are superposition of waves
of opposite circular polarizations.

For $\bf k \parallel \bf B$, the mode $\textrm{a}^{(2)}_\mu$ becomes
pure electric and longitudinal with
$\textbf{e}^{(2)}=\textbf{n}_\parallel$(and also non physical),
whereas $\textrm{a}^{(1)}_\mu$ is transverse
$\textbf{e}^{(1)}=\textbf{n}_\perp$, as it is $\textbf{e}^{(3)}$
 \cite{shabad1}, \cite{shabad2}. In that case $\kappa_{(1)}=
\kappa_{(3)}$, and the circular polarization unit vectors
$(\textbf{e}^{{1}}\pm i \textbf{e}^{{3}})/\sqrt{2}$ are common
eigenvectors of $\Pi_{i j}$ and of the rotation generator matrix
$A^{3ij}$. The quasi-photon moving parallel to $B$ moves with the
speed of light $c$.

The photon magnetic moment is identified as a 3-d vector  which
contains the contribution from electron-positron pairs whose spin
components are oriented along $\textbf{B}$. This is in
correspondence with the fact that after solving the dispersion
equations the resulting entity embodies properties of both free
photons plus electron-positron pairs, leading to a composite
particle which we  name quasi-photons. For them, the phase and group
velocities perpendicular to $B$ are smaller than unity, although,
due to gauge invariance, quasi-photons are massless.  The $e^{\pm}$
virtual pairs contributing to the photon self-energy, if considered
as two-particle states, must have total spin $S=1$, with projections
$S_i =1,-1, 0$. For the quasi-photon propagating along
$\textbf{k}(\perp \textbf{B})$, its observable projection is only
$S_3=0$.

We would like to stress here that the problem of a photon
interacting with the electron-positron field in a magnetic field
implicitly assumes  a Lagrangian describing the interaction of the
electron-positron fields in presence of an external constant
magnetic field (Furry representation). From the Lagrangian a total
Hamiltonian can be written in principle as $H_T=H_{e,B}
+H_{int}+H_{fph}$, where $H_{e,B}, H_{int}, H_{fph}$ are
respectively the Hamiltonian operator for the electrons in the
external field, the interaction Hamiltonian of bounded electrons and
positrons with the photon, and the free photon Hamiltonian (some
gauge condition is assumed to be imposed). Obviously, only
components $P_3$, $J_3$ of the total momentum $P$ and angular
momentum $J_3$ commute with $H_T$ and are conserved along with the
total energy. This is consequence of basic principles in quantum
theory.

 The dispersion equations, obtained as the zeros of
the photon inverse Green function $D^{-1}_{\mu\nu}=0, $
 after diagonalizing the polarization operator
 $\Pi_{\mu\nu}(z_1,z_2,B)$, are
\begin{equation}
k^2=\kappa_{i}(z_2,z_1,B) \hspace{1cm} i=1,2,3.
\end{equation}
 After solving the dispersion equations for $z_1$ in terms of $z_2$ we
get
\begin{equation}
\omega^{(i)2}=\vert\textbf{k}\vert^2+\mathfrak{M}^{2(i)}\left(z_2,
B\right) \label{eg2}
\end{equation}

The refraction index $n^{(i)}$ can be defined as
\begin{equation}
n^{(i)}
=\frac{|\textbf{k}|}{\omega}=(1+\frac{\mathfrak{M}^{2(i)}}{\omega^2})^{-1/2}
\label{refr}
\end{equation}
being different for each eigenmode, leading to the phenomenon of
birefringence. The propagation of light in magnetized vacuum is thus
similar to that in an anisotropic medium.  Gauge invariance
 implies that $\kappa_{(i)}(0,0, B)=0$ and $\mathfrak{M}^{2(i)}(0,B)=0$.
Thus, for parallel propagation $n^{(i)}=1$. By differentiating
(\ref{eg2}) with regard to $B$ we get the relation
$\mu_{\gamma}^{(i)}=-\frac{1}{2\omega}\frac{\partial
\mathfrak{M}^{2(i)}}{\partial B}$, and in consequence
$\mathfrak{M}^{2(i)}=-2\int_0^B \omega \mu_{\gamma}^{(i)}(B') dB' +
f(z_2)$. The function $f(z_2)$ is zero if the series expansion of
$\kappa_{(i)}$ in powers of $B$ is taken as linear in $z_1,z_2$ (see
below). From (\ref{eg2}), we have the approximate dispersion
equation
\begin{equation}
\omega= |\textbf{k}| -\int_0^B \mu^{(i)}_{\gamma}
(z_2,B',|\textbf{k}|)dB'.
\end{equation}
For nonparallel propagation the fact that $n^{(i)}>1$ is a
consequence of photon paramagnetism.

\section{Expressions for the photon anomalous magnetic moment}

We will write now the explicit expressions for the photon magnetic
moment found in \cite{EliPRD}. It was shown that in the regions
$-z_1,z_2 \leq 4m^2$ and $0<B\leq B_c$, the photon is paramagnetic,
since $\mu_{\gamma}^{2,3}>0$. In that small frequencies and weak
field limit the expressions for $\kappa_{i}$ were expanded in series
of even powers of $eB$ and in all positive powers of $z_1, z_2$. We
will take only the linear approximation in $z_1, z_2$. By taking the
first two terms in the $\kappa_{i}^{(0)}$
 series expansion, the magnetic moments are,
\begin{equation}
\mu_{\gamma}^{2}= \frac{14 \alpha z_2}{45 \pi B_c |\textbf{k}|}
\left(b -\frac{52 b^3}{49}\right)>0,
\end{equation}
 and
\begin{equation}
 \mu_{\gamma}^{3}= \frac{8 \alpha z_2}{45 \pi B_c |\textbf{k}|}
\left(b -\frac{24 b^3}{7}\right)>0.
\end{equation}

\subsection{High frequencies  and strong fields case  }

In \cite{EliPRD} was pointed out that in the case $m^2\lesssim
-z_1\leq 4m^2 ,B\lesssim B_c$, the energy gap between successive
Landau energy levels for electrons and positrons is of order close
to the electron rest energy. The photon self-energy diverges for
values of $-z_1= k_{\perp}^{\prime 2}$.

 In the vicinity of the first threshold $n=n^{\prime}=0$ and by
considering $k_\perp\neq0$ and $k_\parallel\neq0$, according to
\cite{shabad1}, \cite{shabad2} the physical eigenwaves are described
by the second and third modes, but only the second mode has a
singular behavior near the threshold. The photon magnetic moment
found in \cite{EliPRD} is
\begin{equation}
\mu_\gamma^{(2)}=\frac{\left(1+\frac{z_2}{2eB}\right)(4m^2+z_1)\alpha
m^3e^{-\frac{z_2}{2eB}}}{\omega B_c \left((4m^2+z_1)^{3/2}+b\alpha
m^3e^{-\frac{z_2}{2eB}}\right)}>0,\label{FRR1}
\end{equation}

 This solution is valid in the
vicinity of the first threshold and agrees with the previously
mentioned numerical result in its paramagnetic property, which is
valid throughout the whole region of transparency (below the first
pair creation threshold).

The expression (\ref{FRR1}) has a maximum near the threshold
\cite{EliPRD}, $z_2 \simeq k_\perp^{\prime 2}$.  By calling
 \begin{equation}\label{mumax}
 m_{\gamma}=\omega (k_\perp^{\prime
2})=\sqrt{4m^2-m^2[2\alpha b\exp(-\frac{2}{b})]^{2/3}}
\end{equation}
 that maximum is $\mu_{\gamma}^{(2)}=\frac{e(1+2b)}{3m_\gamma
 b}\left[2\alpha
b\exp\left(-\frac{2}{b}\right)\right]^{2/3}$.
Numerically for $b
\sim 1$,  $ \mu_\gamma^{(2)}\approx 3\mu^\prime
\left(\frac{1}{2\alpha}\right)^{1/3}\approx 12.85\mu^\prime$.

At this point the authors want to refer again to  paper \cite{Selym}
dealing with topics close to the content of \cite{EliPRD}, but
restricted to asymptotically large magnetic fields. In that paper it
is imposed a linear dependence of the photon magnetic moment with
regard to the photon spin in analogy to the non-relativistic
electron, which is  not true at the relativistic level, and it is
claimed the existence of a non-zero photon magnetic moment component
orthogonal to $B$ . This also cannot be justified a priori since the
symmetry is broken along $B$, and a direction orthogonal to it is
not related to any observable. For instance, $J_1$, $J_2$ did not
commute with the total Hamiltonian $H_T$. Such alleged
``perpendicular component", is shown in the paper \cite{Selym} to
come from the scalar product of two orthogonal vectors,  and in
consequence \textit{\textit{it is recognized as giving a zero
contribution to the $B$ dependent part of $\omega$}}. Thus, that
``perpendicular component" \textit{comes from differentiating zero
with regard to $\textbf{B}$}, which is  mathematically nonsense. It
is also argued to be obtainable by an independent procedure, which a
simple analysis shows that it actually involves again
differentiating  zero with regard to $\textbf{B}$. Based on these
results, it is claimed some  precession of the magnetic moment
around $B$ which has no basis at all (See the Appendix).

\section{ The absorptive region}

Beyond the first threshold $n=n'=0$, for frequencies such that $-z_1
>4m^2$ starts the so-called region of absorption, i.e., for
$4m^2+z_1<0$, $\kappa_{(2)}$ becomes complex, its imaginary part
leading to complex frequencies $\omega + i \Gamma$ after solving the
dispersion equations (the thresholds for absorption would be
slightly lower if the pair resulting from the photon decay forms a
bound state, or positronium. (See \cite{shabad3}, \cite{Usov}). The
quantity $\Gamma$ is finite on the photon mass shell \cite{shabad2}
and it accounts for the probability of photon decay in
electron-positron pairs (the same occurs for higher thresholds).
Thus, in this region the photon magnetic moment cannot be considered
independently of the created electron-positron pairs.

The absorptive region is the continuation of the large frequency,
large magnetic field limit to the region  $-z_1\geq 4m^2$ and fields
$B \gtrsim B_c$. That region is also interesting in astrophysics,
and in cosmology (stars having fields $B \gtrsim B_c$  and early
universe). It is interesting, however, to remark some of the new
features. For instance, although larger values are expected for the
photon magnetic moment than in the region of transparency,  a
negative peak is found for the first threshold of the third mode.
This has no absolute meaning since in that region photons coexist
with electron-positron pairs (the photon has some nonzero
probability of decaying  in pairs or either in positronium) and the
magnetic moment of the created electron-positron pairs (or
positronium) must be added to that of photons. This is point is not
taken into account in \cite{Selym}, where the photon magnetic moment
in supercritical magnetic fields is discussed only on real $\omega$.

The magnetized vacuum background is no longer satisfactorily
described by the Euler-Heisenberg expression $\Omega_{EH}$, and
radiative corrections containing the photon self-energy must be
taken into account. These corrections can be written in the two
equivalent forms \cite{Fradkin} $\Omega_{EH}^{1}=\int_0^e
Tr(de'/e')\int d^4 p G(p)\Sigma(p)=\int_0^e (de'/e')\int d^4 k
D_{\nu\mu}(k)\Pi_{\mu\nu}(k)$,
 where $G(p)$, $\Sigma(p)$ are the
electron Green function and self-energy, respectively. Integration
over $p_{1,2}$ must be understood as sum over Landau numbers $n,n'$.
Appropriate counterterms must be subtracted to make these integrals
convergent. By starting from the second integral, one can write
\begin{equation}\label{OmegaRad}
\Omega_{EH}^{R}=- \frac{1}{(2\pi)^4}\sum_i \int_0^e (de'/e')\int
d^{3}k d\omega \frac{\kappa_{i}}{k^2-\kappa_{i}}
\end{equation}

The leading divergences are eliminated after integrating over $e'$.
For some ranges of $k$, $\omega$ and $B$, $\kappa_{i}$ are divergent
at the pair creation thresholds, getting some imaginary part and
becoming complex, as discussed previously, and this suggests that
quantum vacuum modes at these frequencies become unstable and decay
for fields $B \gtrsim B_c$, in a similar way as observable photons.

\begin{figure}[!htbp]
\includegraphics[width=9cm]{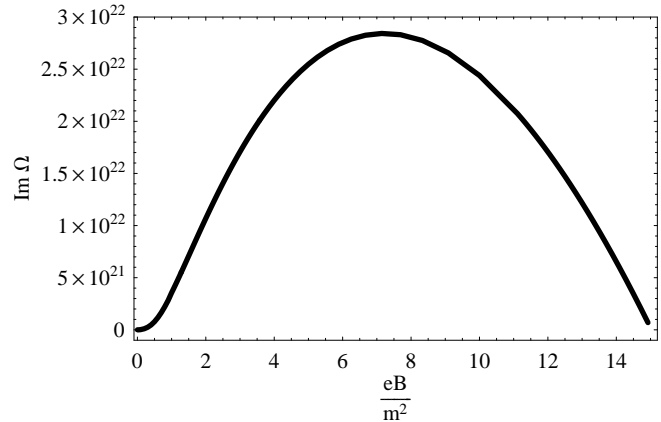}
\caption{\label{ImOm} Imaginary part of vacuum energy.}
\end{figure}

Let's consider in (\ref{OmegaRad}) only the second mode contribution
coming from the first threshold for pair creation $n=n'=0$:
\begin{eqnarray}\label{OmegaRad2}\nonumber
\Omega_{EH 2}^{R}&=& -\frac{1}{(2\pi)^4}\int_0^e (de'/e')\int d^{3}k
d\omega
\frac{\kappa_{2}}{k^2-\kappa_{2}} \\
&=& \frac{1}{(2\pi)^4}\int d^{3}k d\omega \ln
\frac{k^2-\kappa_{2}}{k^2}.
\end{eqnarray}
From (\ref{OmegaRad2}) it is easy to obtain an expression for the
imaginary part. We assume that $Re  \kappa_{2}$ is small in the
absorptive region, where the main contribution comes from $Im
\kappa_{2}$. Thus, in the first line of (\ref{OmegaRad2}) the
quantity
\begin{eqnarray}\label{OmegaRad3}\nonumber
\frac{\kappa_{2}}{k^2-\kappa_{2}}&=&\frac{i Im\kappa_{2}}{k^2-i Im
\kappa_{2}}\\
&=&\frac{-Im\kappa_{2}^2 + i  k^2 Im \kappa_{2} }{k^4 + Im
\kappa_{2}^2}
\end{eqnarray}
From (\ref{OmegaRad3})
\begin{equation}\label{ImOmegaRad2}
\mathrm{Im}  \Omega_{EH 2}^{R}= \frac{1}{(2\pi)^4}\int d^{3}k
d\omega \arctan -\frac{\mathrm{Im} \kappa_{2}}{k^2},
\end{equation}
where (see \cite{shabad2})
\begin{equation}\label{ImKappa2}
  \mathrm{Im} \kappa_{2}=\frac{4\alpha eB m^2 \Theta (-4m^2-z_1)}{\sqrt{z_1(z_1+4m^2)}}
   \exp\left(-\frac{z_2}{2eB}\right),
\end{equation}
and $\Theta (-4m^2-z_1)$ is the step function. The integration over
$z_1$ is extended along the half-plane $z_1 \leq -4m^2$, and  for
$z_2$, in the half-plane $z_2 \geq 0$. The numerical calculation of
(\ref{ImOmegaRad2}) is shown in fig. \ref{ImOm}.

By observing the figure we see that for fields $2B_c<B<12 B_c$ we
have ${Im} \kappa_{2}\sim 10^{22}$ in energy units. This would
correspond to $\sim 10^{28}$ electron-positron pairs or photons,
after the pairs decay. Magnetized vacuum becomes unstable and may
decay finally in electromagnetic radiation through that mechanism,
for fields $B \geq B_c$. The question immediately comes about the
source of matter or energy to account for the created $e^{\pm}$
pairs or photons. We conclude that the source of the field would
contribute with a compensating amount of energy. For instance, if
the field is produced by a star through the dynamo effect, the
vacuum decay would be produced at the expense of a small slowing
down of the rotating energy, and a consequent decrease of  the
magnetic field intensity. Magnetic energy would be transformed into
radiation beams, propagating along $B$. Thus, the vacuum decay
through the present mechanism might play a role in the $\gamma$-ray
emission of strongly magnetized stars.

We recall also that at present QED in unstable vacuum (see for
instance \cite{Grib}, \cite{Fradkin2} and its references) has
growing interest, and the possibility of observing vacuum decay in
critical electric fields in terrestrial laboratories (see
 \cite{Rafelski}) is becoming realistic thanks to the development of
high power pulse lasers technology.

\section{Acknowledgments}
The authors thank thanks W. Greiner and A.E. Shabad for several
comments and important remarks. They thank also OEA-ICTP for support
under Net-35. One of the authors (H.P.R.) thank Cesar Zen
Vasconcellos and Jorge Horvath for hospitality at the International
Workshop on Astronomy and Reltivistic Astrophysics (IWARA) 2009. The
present paper is an updated version of a talk given on it.

\section{Appendix}
We are assuming throughout this paper the constancy of the external
magnetic field (as also is done  in \cite{Selym}). If $B$ varies, we
are dealing with a different problem. The ``demonstration" made in
the Appendix of \cite{Selym} is due to a mistake. It is equivalent
to the following procedure. Let us start from a rotated system of
coordinates where $B_y$ and $B_z$ are the components (of the spatial
part) of the magnetic field tensor $F_{ij}$. We have

\begin{equation}\label{F}
F_{ij}k_j=((B_zk_y-B_yk_z), -B_zk_x, B_yk_x)
\end{equation}
Its square is,
\begin{equation}\label{kFk}
   (kF^2k)=k_x^2(B_z^2+B_y^2)+(B_zk_y-B_yk_z)^2=[{\bf B}\times {\bf
   k}]^2.
\end{equation} The contribution of $\mu_{y}=(\partial \omega/\partial
kF^2k)(\partial kF^2k/\partial B_y$ to the
perpendicular-to-$\textbf{B}$ magnetic moment would be proportional
to
\begin{equation}\label{18}
   \frac{\partial kF^2k}{\partial
   B_y}=2k_x^2B_y-2(B_zk_y-B_yk_z)k_z.
\end{equation} If we assume that the limit $B_y\rightarrow 0$ can be taken, it would lead to

\begin{equation}\label{B_y=0}
   \left.\frac{\partial kF^2k}{\partial
   B_y}\right|_{B_y=0}=-2|{\bf B}|k_yk_z.
\end{equation}
Since $B_z=|{\bf B}|$, when $B_y=0$, at first sight it seems that a
perpendicular component to the magnetic moment arises, depending on
the transversal, $k_y$, and longitudinal, $k_z$, momenta.

Let us show that this demonstration has a mistake, due to the fact
that it bypasses a constraint implicit in the vector product
definition, since $\textbf{B} \times \textbf{k}=\textbf{B} \times
[\textbf{k}_{\parallel}+\textbf{k}_{\perp}]=\textbf{B} \times
\textbf{k}_{\perp}$, and it implies $\textbf{B} \times
(\textbf{k}_{\parallel})=0$.

\begin{figure}[!htbp]
\includegraphics[width=7cm]{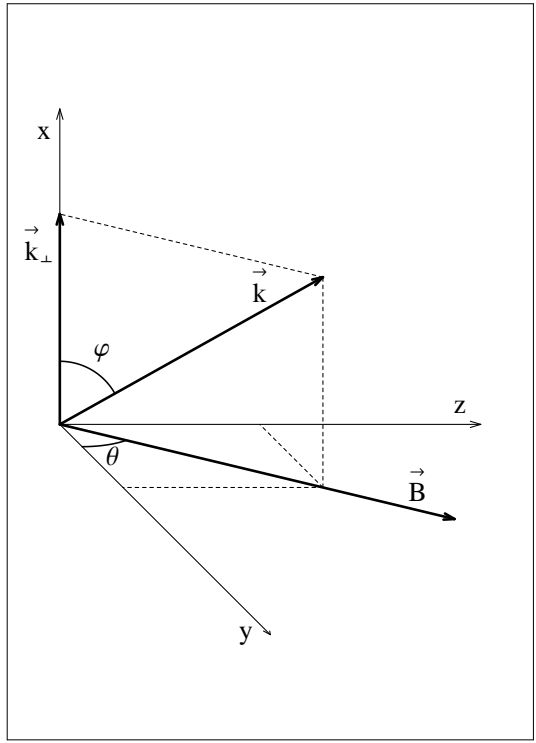}
\caption{\label{last} Relation between components of the vector
product $\textbf{B}\times\textbf{k}$.}
\end{figure}

First of all let us consider the two sides of equation (\ref{kFk}).
From the general definition we expect that $[{\bf B}\times {\bf
   k}]^2=B^2 k^2_{\perp}=(B_y^2 + B_z^2)k^2_{\perp}$. Thus, by
   comparison with the left hand side, it must be either
   $B_zk_y-B_yk_z=0$ and $k_{\perp}=k_{x}$, or  $B_zk_y-B_yk_z= B k'_{\perp}$, must hold after some algebra, where
   $k'_{\perp}$ is some other component of $\textbf{k}$ orthogonal to
   $\textbf{B}$.  Let us assume that the
plane ${\textbf{B,k}}$ is chosen  as orthogonal to the plane
$(y,z)$. This selection of the coordinate system can always be done
and in it only the component $\textbf{k}_{\parallel}$ lies on the
$y,z$ plane. Then $\textbf{k}_{\parallel}=\textbf{k}\sin \varphi$
(where $\varphi$ is the polar angle of $\textbf{k}$ with regard to
$x$-axis) has components $k_y =k_{\parallel} \cos \theta,
k_z=k_{\parallel} \sin \theta$ ($\theta$ is the azimuthal angle) and
in consequence in (\ref{kFk}), (\ref{18}) $B_y k_z -B_z k_y =Bk \sin
\varphi(\cos \theta \sin \theta-\sin \theta \cos\theta)=0$
identically, and $\partial (B_y k_z -B_z k_y)/\partial B_y=0 $,
since de derivative of zero with regard to any quantity is zero. See
fig. \ref{last}
We will refer to a mechanical analogy of the  ``perpendicular
 diamagnetic component" problem :
the case of a point of mass $m$ rotating uniformly around a center
(as a simplified version, for instance, of a planet around the sun;
we may assume the existence of a central force). With regard to some
adequately chosen reference frame, its kinetic energy can be written
as $E=m (\omega \times \textbf{r})^2/2$. The angular velocity
$\omega$ is taken along the 3-rd axis, since the particle rotates in
the 1,2 plane. Here $\omega$ is the analog of $B$. From $E$, we get
the angular momentum, (the analog of the magnetic moment)
\begin{eqnarray}
L&=&\partial E/\partial\omega =m(\textbf{e}_3 \times
\textbf{r})\cdot (\omega \times \textbf{r})\\ \nonumber
&=&\textbf{e}_3 \cdot (m \textbf{r}\times (\omega \times
\textbf{r}))\\ \nonumber &=& m r^2 \omega .
\end{eqnarray}
Where $L=m r^2 \omega$ is the modulus of the angular momentum vector
$\textbf{L}= L \textbf{e}_3$  directed along $\omega$. It would be
nonsense to claim about the existence of an angular momentum
component orthogonal to $\omega$, which would violate the laws of
planetary motion.

\end{document}